\documentclass[12pt]{article}
\usepackage{epsfig}

\def\be{\begin{equation}}
\def\ee{\end{equation}}
\def\ba{\begin{array}{c}}
\def\ea{\end{array}}
\def\p{\partial}
\def\ben{$$}
\def\een{$$}
\begin{document}

\titlepage

 \begin{center}{\Large \bf
Quantum toboggans with  two branch points
 }\end{center}

\vspace{5mm}

 \begin{center}
Miloslav Znojil

\vspace{3mm}

\'{U}stav jadern\'e fyziky\footnote{e-mail: znojil@ujf.cas.cz }
 AV \v{C}R, 250 68 \v{R}e\v{z}, Czech
Republic

\end{center}

\vspace{5mm}

\section*{Abstract}

In an innovated version of ${\cal PT}-$symmetric Quantum
Mechanics, wave functions $\psi^{(QT)}(x)$ describing quantum
toboggans (QT) are defined along complex contours of coordinates
$x(s)$ which spiral around the branch points $x^{(BP)}$. In the
first nontrivial case with $x^{(BP)}=\pm 1$, a classification is
found in terms of certain ``winding descriptors" $\varrho$. For
{\em some} $\varrho_0$, a mapping $x^{(\varrho)}(s) \to
y^{(0)}(s)$ is presented which rectifies the contours and which
enables us to extend, to our QTs, the standard proofs of the
reality/observability of the energy spectrum.

 \vspace{9mm}

\noindent
 PACS 03.65.Ge


 \begin{center}
\end{center}

\newpage

\section{Introduction \label{s1} }

The current one-dimensional Schr\"{o}dinger equation
 \be
 -\frac{\hbar^2}{2m}\,\frac{d^2}{dx^2} \,\psi_n (x) + V(x)
  \,\psi_n (x)= E_n \,\psi_n (x)\,
   \label{SE}
 \ee
for bound states $\psi_n(x)$  and for their energies $E_n$
contains, very often, a  phenomenological potential $V(x)$ which
is a holomorphic function in a complex domain of $x$. It is well
known to mathematicians \cite{Alvarez,Caliceti,BG} that this
property is shared by the two linearly independent solutions
$\psi^{(special)}_{(\pm)}(x)$ of eq.~(\ref{SE}) and also by their
arbitrary superpositions
 \be
 \psi^{(general)}(x)=c_+\psi^{(special)}_{(+)}(x)+
 c_-\psi^{(special)}_{(-)}(x)\,.
 \label{wavefs}
 \ee
In 1993, Bender and Turbiner \cite{BT} communicated a few
interesting consequences to the physics community. They stressed
that the Schr\"{o}dinger's differential eq.~(\ref{SE}) can be
perceived as comprising {\em several} eigenvalue problems at once,
depending on our choice of the (in principle, non-equivalent)
asymptotic boundary conditions {\em in the complex plane of $x$}.
Five years later, Bender and Boettcher presented a much better and
more explicit formulation of this idea in their famous letter
\cite{BB} which initiated the subsequent quick development of the
whole new branch of Quantum Theory nicknamed ${\cal PT}-$symmetric
Quantum Mechanics \cite{BBjmp,Carl}. In this framework, it is now
agreed (cf. also the short summary of the state of the art in
section \ref{s2} below) that a phenomenological Schr\"{o}dinger
eq.~(\ref{SE}) can be defined not only along the current real line
of $x$ but also, in its various analytically continued forms,
along a suitable ${\cal PT}-$symmetric (which means left-right
symmetric) complex integration path $x=x(s)$ with $s \in
(-\infty,\infty)$.

In our recent letter \cite{I} we extended slightly the scope of
such an innovative approach to eq.~(\ref{SE}). In essence, we
imagined that the holomorphy domain ${\cal S}$ of an analytic wave
function (\ref{wavefs}) with a branch point (say, in the origin,
$x^{(BP)}=0$) can be topologically nontrivial. In this way, also
the integration contour $x(s)$ pertaining to eq.~(\ref{SE}) can be
continued over {\em several} sheets ${\cal R}_0$, ${\cal R}_{\pm
1}$, \ldots of the complete Riemann surface ${\cal R}$ of
$\psi(x)$. We argued (cf. also section \ref{s3} below) that one
can construct the topologically nontrivial, ``quantum-tobogganic"
(QT1) integration path $x^{(QT)}(s)$ which connects a pair of its
asymptotes only after having made an $N-$tuple turn around the
singularity in the origin. In our subsequent, more technical paper
\cite{II} we expressed an opinion that in the next,
``double-spire" (QT2) case, the problem of quantum toboggans
``does not seem to admit closed-form solutions".

The latter remark proved over-skeptical and we intend to disprove
it in what follows. Firstly we shall show how one can replace the
QT1-related winding number $N$ by an appropriate QT2-related
winding descriptor $\varrho$ (cf. section \ref{s6}). Next we point
out that one can extend at least some of the analytic and
constructive QT1-related considerations {\em immediately} to the
two-spired QT2 context. This is supported by section \ref{s4}
where, for a subset of descriptors $\varrho_0$, the necessary QT2
rectification mapping $x^{(\varrho)}(s) \to y^{(0)}(s)$ is found
and discussed. In particular, we show there that in full analogy
with the single-spire case, the rectified, ``effective" potential
becomes ${\cal PT}-$symmetric and, in spite of its slightly more
involved explicit form, amenable to the standard treatment,
therefore.

In the summary section \ref{s5} we re-emphasize that in a complete
parallel to the non-tobogganic  ${\cal PT}-$symmetric quantum
models, the practical applicability of their QT innovations will
crucially depend upon the availability of the individual rigorous
proofs of the reality/observability of their energies. In this
sense, our present constructive demonstration of the closed-form
equivalence between {\em some} tobogganic and non-tobogganic paths
can be perceived as a promising first success in this direction.

\section{Models with complex coordinates \label{s2} }

\subsection{Analytically continued wave functions \label{s2a} }

The principle of correspondence translates the one-dimensional
motion of a {\em classical} particle  into its quantum parallel in
which the Hamiltonian $H =-\p^2_x+ V(x)$ (written in units $\hbar
= 2m = 1$) is an operator in ${\cal L}_2(I\!\!R)$ and in which the
real spectrum represents the observable energies. In this context,
Bender and Turbiner \cite{BT} contemplated the harmonic-oscillator
Hamiltonian $H^{(HO)} =-\p^2_x+ x^2$ and noticed that a
replacement of the real axis of $x \in I\!\!R$ by the imaginary
axis of $x \in {\rm i}\,I\!\!R$ preserves the reality of the
energies. Later, Bender and Boettcher \cite{BB} noticed that the
harmonic-oscillator spectrum remains real {\em and} bounded below
after a ${\cal PT}-$symmetric complexification of the standard
boundary conditions (note that the operator ${\cal P}$ represents
parity while an antilinear complex conjugation ${\cal T}$ mimics
time reversal). They argued that with the choice of $V(x)=x^2$ and
$x = y-{\rm i}\,\varepsilon \notin I\!\!R$ in eq.~(\ref{SE}) where
$\varepsilon> 0$ and $y \in I\!\!R$, one arrives at an {\em
equivalent} Schr\"{o}dinger bound-state problem with a different,
{\em manifestly non-Hermitian} potential,
 \be
 \left ( -\frac{d^2}{dy^2}+ y^2 -2\,{\rm i}\,\varepsilon\,y
 \right )\,\tilde{\psi}_n(y) = \tilde{E}_n \,\tilde{\psi}_n(y)\,.
 \label{SEHO}
 \ee
Still, the new energies remained {\em all} safely real due to the
identity $\tilde{E}_n =E_n+\varepsilon^2$. The asymptotics of wave
functions $\psi^{(HO)} (x)= \exp(-x^2/2) \times {\rm polynomial}$
vanishing at the real $x\to \pm \infty$ remained vanishing also in
the asymptotic domain of $y\to \pm \infty$ since
$\tilde{\psi}_n(y) =\psi_n(y-{\rm i}\,\varepsilon)$.

\subsection{Spiked and ${\cal PT}-$symmetric harmonic
oscillator
 \label{s2b} }

In the language of ref.~\cite{BB} the analytically continued wave
function ``lives" on a ${\cal PT}-$symmetric integration curve of
$x$. Its above-mentioned HO example is just a very special case of
a broad class of the manifestly non-Hermitian ${\cal
PT}-$symmetric Hamiltonians with the real spectrum \cite{Carl}.

The spiked harmonic oscillator of ref.~\cite{ptho} can be selected
as another, slightly less elementary element of the family. The
one-dimensional Hamiltonian $H^{(HO)}$ of paragraph \ref{s2a} is
replaced by its $D-$dimensional (or rather ``radial" or ``spiked")
version, possessing a centrifugal-like pole in its potential term.
After the same complexification of the coordinate $x=y-{\rm
i}\,\varepsilon$ as above, a merely slightly more sophisticated
Schr\"{o}dinger equation results,
 \be
 \left ( -\frac{d^2}{dy^2}+ y^2 -2\,{\rm i}\,\varepsilon\,y
 +\frac{\alpha^2-1/4}{y^2 -2\,{\rm
 i}\,\varepsilon\,y-\varepsilon^2}
 \right )\,\tilde{\psi}_n^{(\alpha)}(y) = \tilde{E}_n^{(\alpha)}
  \,\tilde{\psi}_n^{(\alpha)}(y)\,.
 \label{SEHOspi}
 \ee
It is exactly solvable again, with $\psi^{(\pm |\alpha)|}_n (x)=
x^{1/2\pm |\alpha|}\,\exp(-x^2/2) \times {\rm a\,polynomial}$. As
a new feature of the model, a branch point is encountered in the
wave functions at $x^{(BP)}= y^{(BP)}-{\rm i}\,\varepsilon=0$.

In \cite{ptho} the holomorphy domain ${\cal S}$ has been
restricted to the complex plane with an upwards-running cut.
In contrast to the non-spiked case (where the spectrum has not
been influenced by the shift  $\varepsilon$ at all), the presence
of the branch point singularity introduced a difference between
the spectra at $\varepsilon=0$ (where $x \in (0,\infty)$) and at
$\varepsilon> 0$ (with twice as many levels $\tilde{E}_n^{(\pm
|\alpha| )} =4n+2\pm 2\,|\alpha| +\varepsilon^2$  \cite{ptho}).

\subsection{A less usual, nonstandard harmonic oscillator
 \label{s2c} }

Common wisdom tells us that the complexified contours of the HO
coordinates $x(s)$ can only be introduced as a
left-right-symmetric smooth deformation of the real axis which
stays inside the correct asymptotic wedges,
 \be
  x(s) = \pm |\,x(s)|\, e^{{\rm i}\,\xi(s)}\,,\ \ {\rm where}\
  \xi(s) \in \left (-\frac{\pi}{4},\frac{\pi}{4}\right )
  \ {\rm and}\ |\,x(s)|  \gg 1\,
  \ {\rm for}\ |\,s|  \gg 1\,
 \,.
 \label{Sshift}
 \ee
For all the other asymptotically quadratic potentials the recipe
is believed to remain the same. The presence of the branch points
and cuts merely seems to force us to restrict the freedom of
choosing $x(s)$ at the smaller $|s|$.

In ref.~\cite{I} we revealed that an alternative, U-shaped
integration contour could be used as well (cf. Figure Nr.~5 in
{\it loc. cit.}). Thus, in principle, one is allowed to integrate
Schr\"{o}dinger eq.~(\ref{SEHOspi}) along a contour
$x(s)=y(s)-{\rm i}\,\varepsilon$
where, in the asymptotic domain, one chooses $\xi(s) \in
(-\pi/8,\pi/8)$ and sets, say,
 \be
   y^{(U)}(s) =  x^{(U)}(s)+{\rm i}\,\varepsilon =\,
 \left \{
 \begin{array}{ll}
   |\,s|\, e^{-11\,{\rm i}\,\pi/8}
  \exp[{\rm i}\,\xi(s)],\,&\ \ s \ll -1\,,\\
  |\,s|\, e^{3{\rm i}\,\pi/8}
  \exp[{\rm i}\,\xi(s)],&\ \ s \gg 1\,.
 \ea
 \right  .
 \label{Supshift}
 \ee
This leads to the anomalous boundary conditions imposed upon
$\tilde{\psi}_n(y) =\psi_n(x)$,
 \be
  \lim_{s\to \pm, \infty}\, \psi\left[\,x^{(U)}(s)\right ] = 0\,.
 \label{SEHOanobc}
 \ee
It is worth noting that the wave functions then acquire an
interesting, {less usual} asymptotic form. Indeed, after one
modifies slightly the recipe of ref.~\cite{ptho} and after one
re-arranges eq.~(\ref{SEHOspi}) as a confluent hypergeometric
differential equation, it is easy to deduce, in an exercise left
to the reader, that the anomalous boundary
conditions~(\ref{SEHOanobc}) lead to the untilded wave functions
in the apparently anomalous explicit form $\psi^{(\pm |\alpha)|}_n
(x)= x^{1/2\pm |\alpha|}\,\exp(+x^2/2) \times {\rm
a\,polynomial}$.

\section{QT models with the single branch point   \label{s3} }

\subsection{A tobogganic alternative to contour
(\ref{Supshift})
 \label{s2d}}

The introduction of the concept of quantum toboggans (QT,
\cite{I}) can be perceived as a mere slight innovation of the
traditional quantization recipes, with the boundary conditions
$\lim_{s\to \pm \infty}\psi^{(QT)}[x(s)]=0$ located on the {\em
different} Riemann sheets of $\psi^{(QT)}(x)$. Such a definition
does not necessarily imply an increase of technical complications.
This can be easily illustrated on the example of paragraph
\ref{s2c} where the set of the U-shaped contours (\ref{Supshift})
(located, conveniently, on the zeroth Riemann sheet ${\cal R}_0$
of $\psi(x)$) can be analytically continued, without any changes
in the spectrum, to the tobogganic contours where the upper and
lower asymptotes of the curve $x(s)$ or $y(s)$ may be understood
as lying on the first and minus first Riemann sheet ${\cal R}_1$
and ${\cal R}_{-1}$, respectively, with any $\xi \in
(-\pi/8,\pi/8)$ and with some auxiliary small $\delta>0$ in
 \be
   x^{(N=1)}(s) =y^{(N=1)}(s)-{\rm i}\,\varepsilon =\,
 \left \{
 \begin{array}{ll}
   |\,s-\eta|\, e^{-13\,{\rm i}\,\pi/8}
  \exp[{\rm i}\,\xi(s)]+{\rm i}\,\delta,\,&\ \ s \ll -1\,,\\
  |\,s-\eta|\, e^{5{\rm i}\,\pi/8}
  \exp[{\rm i}\,\xi(s)]+{\rm i}\,\delta,&\ \ s \gg 1\,
 \ea
 \right  .
 \label{Uptobshift}
 \ee
where $\eta\gg 1$ is the value of $s$ at which both the branches
of curve $x^{(N=1)}(s)$ return to the zeroth Riemann sheet ${\cal
R}_0$ in a way sampled, say, by Figure Nr.~3 in ref.~\cite{I}.

\subsection{A rectification transformation of QT eq.~(\ref{SEHOspi})
+ (\ref{Uptobshift}) \label{uho} }

We saw that the harmonic-oscillator differential equation
(\ref{SEHOspi}) offers one of the simplest illustrations of the
concept of quantum toboggan with the single branch point and with
the first nontrivial winding number $N = 1$ in
eq.~(\ref{Uptobshift}). Of course, at a generic, irrational real
exponent $\alpha > 0$ in eq.~(\ref{SEHOspi}), the Riemann surface
${\cal R}$ of all the wave functions is composed of infinitely
many sheets ${\cal R}_{\pm k}$. This means that in a way discussed
thoroughly in ref.~\cite{I}, one may generalize
eq.~(\ref{Uptobshift}) and construct the HO QT curves with any
integer winding number $N$.

In an opposite direction, our knowledge of $\psi^{(\pm \alpha)}
(x) \sim x^{1/2\pm \alpha}$ near the origin inspires us to perform
the following elementary change of the variables $x=y-{\rm
i}\,\varepsilon \to z$ such that
 \be
 {\rm i}\,x=({\rm i}\,z)^2\,,\ \ \ \ \ \
 \psi_n(x)=\sqrt{z}\,\varphi_n(z)\,.
 \label{rectifi1}
 \ee
It is easy to check that this transforms our tobogganic
Schr\"{o}dinger eq.~(\ref{SEHOspi}) into a  {\em strictly
equivalent} differential-equation problem
 \be
  \left ( -\frac{d^2}{dz^2}+ 4\,z^6
 +4\,E_n\,z^2 +
 \frac{4\,\alpha^2-1/4}{z^2}
 \right )\,\,\varphi(z) = 0\,
 \label{SEHOspiprime}
 \ee
which is to be read as another Schr\"{o}dinger equation which must
be considered at the strictly vanishing energy.

It is important to notice that once we drop all the (by
construction, inessential) corrections due to $\delta$, our new
bound-state problem (\ref{SEHOspiprime}) is defined on the new,
transformed contour of
 \be
   z^{(N=0)}(s) =\,
 \left \{
 \begin{array}{ll}
  \sqrt{ |\,s-\eta|}\, e^{-9\,{\rm i}\,\pi/16}
  \exp[{\rm i}\,\xi(s)/2]+{\cal O}(\delta/\eta),\,&\ \ s \ll -1\,,\\
  \sqrt{|\,s-\eta|}\, e^{{\rm i}\,\pi/16}
  \exp[{\rm i}\,\xi(s)/2]+{\cal O}(\delta/\eta),&\ \ s \gg 1\,
 \ea
 \right  .
 \label{Uptoshit}
 \ee
which only slightly deviates from the straight line and is,
therefore, {\em manifestly non-tobogganic}. In the other words,
our transformation (\ref{rectifi1}) rectified the QT contour and
returned all our considerations to the {\em single} complex plane
of the new coordinate $z$, equipped with the upwards-oriented cut.
The price to be paid for the rectification $[(N=1) \to (N=0)]$
looks reasonable. The new representation (\ref{SEHOspiprime}) of
our toy QT bound-state problem contains the same centrifugal spike
with an enhanced strength. In addition, the new  potential becomes
manifestly level-dependent while its sextic anharmonic form still
remains sufficiently elementary.

It is essential that the rectified equivalent (\ref{SEHOspiprime})
+(\ref{Uptoshit}) of our original QT bound-state problem
(considered at a fixed energy and called, usually, ``Sturmian"
eigenvalue problem) proves {\em manifestly ${\cal PT}-$symmetric}.
This returns us back to the safe territory of the standard theory
\cite{Carl} where the methods of the necessary proof of the
reality of the spectrum are already well known \cite{DDT}. {\it
Vice versa}, the obvious universality as well as an extreme
simplicity of the rectification transformation (\ref{rectifi1})
indicate that in a search for some topologically really nontrivial
QT models one has to move to the systems with at least two branch
points.

\section{An enumeration of the models with the paths $x^{(QT)}(s)$
encircling the two branch points  \label{s6} }

Once we assume the presence of a pair of branch points in
$\psi(x)$, say, at $x^{(BP)}_{(\pm)}=\pm 1$, each curve ${x(s)}$
which leaves its ``left", $s \ll -1$ asymptotic branch will have
to stop (say, below one of the branch points) and, during the
further increase of $s$, it will have to pick up one of the
following four options of

\begin{itemize}

 \item
winding counterclockwise around the left branch point
$x^{(BP)}_{(-)}$ (this option may be marked by the letter $L$),

 \item
winding counterclockwise around the right branch point
$x^{(BP)}_{(+)}$ (marked by the letter $R$),

 \item
winding clockwise around the left branch point $x^{(BP)}_{(-)}$
(marked by the letter or inverse-turn symbol $Q=L^{-1}$),

 \item
winding clockwise around the right branch point $x^{(BP)}_{(+)}$
(marked by $P=R^{-1}$).

\end{itemize}

 \noindent
Using the four-letter alphabet, each individual ${\cal
PT}-$symmetric curve ${x}=x^{(\varrho)}(s)$ can be uniquely
characterized by a word $\varrho$ of an even length $2N$. Once we
ignore the empty symbol $\varrho=\emptyset$ as trivial,
corresponding to the mere non-tobogganic straight line, we shall
encounter precisely four possibilities in the first nontrivial
case with the two turns around branch points,
 \ben
 \varrho \in \left \{LR\,, L^{-1}R^{-1}\,, RL\,,R^{-1}L^{-1}
 \right \}\,,\ \ \ \ \ N=1\,.
 \een
The physical requirement of  ${\cal PT}-$symmetry acquires the
form of the $L \leftrightarrow R$ interchange after the
transposition (i.e., $\varrho \to \varrho^T$ which means reverse
reading) of all the ``admissible" words $\varrho$ of the length
$2N$. This means that we can always decompose each word in two
halves, $\varrho= \Omega \bigcup \Omega^T$. This reduces our
``enumeration" problem to the specification of all the words
$\Omega$ of the length $N$, giving the simplified list of the four
items
 \ben
 \Omega \in \left \{L\,, L^{-1}\,, R\,,R^{-1}
 \right \}\,,\ \ \ \ \ N=1\,,
 \een
at $N=1$, or the dozen of their descendants
 \ben
 \left \{
 LL,
 LR, RL, RR, L^{-1}R, R^{-1}L,
  LR^{-1}\,, RL^{-1}\,,
  L^{-1}L^{-1}, L^{-1}R^{-1}, R^{-1}L^{-1},R^{-1}R^{-1}
 \right \}
 \een
at $N=2$ where, among all the $4^2=16$ eligible combinations of
two letters, the four words, viz, $ L L^{-1}\,, L^{-1}L\,,R
R^{-1}$ and $R^{-1}R$ are ``not allowed" because their ``real
length" [meaning the total winding number of the corresponding
tobogganic ${x}^{( \Omega \cup \Omega^T)}(s)$] is in fact shorter
than two.

The latter observation indicates that at any total winding index
$N$, the necessary determination of all the ``labelling words"
$\Omega$ degenerates to the much easier specification of all the
$N-$letter words $\Omega^{(NA)}$ which are ``not allowed". Thus,
at $N=3$ one can split the $\Omega^{(NA)}$ set in two disjoint (or
conjugate) equal-size subsets $\Omega^{(NAL)} \bigcup
\Omega^{(NAR)}$ with the prevalence of the (possibly, also
inverse) $L$s and $R$s, respectively. Next, each of them (say,
$\Omega^{(NAL)}$) further decomposes in two disjoint,
non-equal-size subsets $\Omega^{(NAL3)} \bigcup \Omega^{(NAL2)}$
containing three or two $L-$type letters, respectively. Now,
obviously, in the first subset $\Omega^{(NAL3)}$ one can have one
or two inversions so that the total number of the eligible words
of this class is six. In the second subset $\Omega^{(NAL2)}$ we
just add an $R-$type letter (i.e., $R$ or $ R^{-1}$) to $ L
L^{-1}$ or $L^{-1}L$ yielding the final eight nonequivalent
possibilities. Altogether, having started from $4^3=64$ words, we
have to cross out all the 28 ``not allowed" ones. The total number
of the three-letter labels $\Omega$ is equal to 36.

In the next step where $N=4$ one finds, {\em mutatis mutandis}, 14
elements in $\Omega^{(NAL4)}$ and 24 elements in
$\Omega^{(NAL3)}$, yielding the subtotal of 76 not allowed words
after $L \leftrightarrow R$ conjugation. Without an explicit use
of the conjugation, the elements of the remaining set
$\Omega^{(NAL2)}$ can be finally listed as belonging to the class
$\Omega^{(NAL21)}$ (of non-allowed words containing a single
inversion, 16 elements), $\Omega^{(NAL22)}$ (containing two
inversions, 8 elements) or $\Omega^{(NAL21)}$ (three inversions,
16 elements). For the four-letter labels $\Omega$ this leads to
their final number equal to $256 - 76 - 40 = 140$.

\section{Rectifiable contours ${x^{(\varrho_0)}(s)}$
and the reality of the spectrum
 \label{s4} }

Our combinational exercise presented in the previous section
\ref{s6} indicates that the total number of the nonequivalent
toboggans will grow very quickly with the growth of the word
length $2N$ of the winding descriptor $\varrho$. This provokes, on
one side, the expectations of a wealth of the bound-state spectra,
all of which would be attributed to the {\em same}
phenomenological potential $V(x)$. At the same time it is not
clear how we shall be able to make a quantitative prediction of
these variations of the observables in dependence on the winding
descriptors $\varrho$. For this reason let us now restrict our
attention to the mere rectifiable contours with characteristics
$\varrho=\varrho_0$.

\subsection{The changes of variables preserving the
pair of the branch points  $x^{(BP)}=\pm 1$ \label{dva2} }

In our forthcoming study of the paths
$x^{(QT)}(s)=x^{(\varrho)}(s)$ which encircle the doublets of
branch points $x^{(BP)}=\pm 1$ we feel inspired by the
non-tobogganic ${\cal PT}-$symmetric models of Sinha and Roy
\cite{Sinha}. They achieved the exact solvability of their sample
equations with more branch points by the Darboux-transformation
technique. The same technique also enabled them to prove the
reality of the spectra.

In the generic, nontrivial, genuine tobogganic cases using the
potentials $V({\rm i}\,x)$ which are not solvable exactly, the
proof of the observability (i.e., of the reality) of the spectrum
will be much more difficult even in the first nontrivial case
where $x^{(BP)}=\pm 1$. Fortunately, one can again try to proceed
in analogy with the above-outlined rectification recipe applied to
the models with the single branch point  $x^{(BP)}=0$~\cite{I,II}.

In advance, let us emphasize that in comparison with the
single-spire cases, we shall only partially succeed. Still, for
{\em some} (i.e., this time, not all) quantum toboggans with
$\varrho=\varrho_0$, we shall again be able to find an equivalence
transformation between our tobogganic Schr\"{o}dinger equation  of
the form
 \be
 \left [
 -\frac{d^2}{dx^2}
 + \frac{\ell(\ell+1)}{(x-1)^2}
 + \frac{\ell(\ell+1)}{(x+1)^2}
 +
 V(ix)
  \right ]
 \,\psi (x)=
 E \,\psi (x)\,.
 \label{inio}
 \ee
and its zero-energy and state-dependent rectified partner
 \be
 \left [
 -\frac{d^2}{dz^2}+
 U_{eff}({\rm i}\,z)
  \right ]
 \,\varphi (z)=0\,.
 \label{fino}
 \ee
In essence, we shall require that eq.~(\ref{fino}) is defined on
the single, zeroth Riemann sheet ${\cal R}_0$ so that it may be
assumed tractable by the standard methods of  ${\cal
PT}-$symmetric Quantum Mechanics \cite{Carl}. Indeed, the latter
problem {\em will be} manifestly ${\cal PT}-$symmetric once we
show that it can be re-written in the form
 \ben
 U_{eff}({\rm i}\,z)=U({\rm i}\,z)+
  \frac{\mu(\mu+1)}{(z-1)^2}
 + \frac{\mu(\mu+1)}{(z+1)^2}
 \ \equiv\ U({\rm i}\,z)+
  2\,\frac{\mu(\mu+1)[1-
  ({\rm i}\,z)^2]}{\left [1+({\rm i}\,z)^2\right ]^2}
  \,.
 \een
On the entirely pragmatic level we believe that  a numerically
robust character of the equivalence mapping between
eqs.(\ref{inio}) and (\ref{fino}) is vital for the preservation of
the reliability of the practical numerical calculations. The
necessity of a non-numerical construction of the equivalence
mapping should be emphasized since it opens the chances of finding
not only the transparent, closed-form ``direct" map but also its
inversion. Such a knowledge would allow us {\em to start} from a
known, non-tobogganic ${\cal PT}-$symmetric problem (\ref{fino})
(selected as possessing the safely real parameters of course) and
{\em to construct}, afterwards, some nontrivial, physical, ``QT2"
models (\ref{inio}) with the real spectrum and, hence, with the
possible practical significance.

\subsection{The rectification of the QT2 contours
at $\varrho=\varrho_0$}

We saw that the toboggans with the single branching point were
much more easy to classify \cite{I}. For {all of them}, in
addition, it was fairly easy to find an elementary change of the
variables in Schr\"{o}dinger equation  which transformed a given
tobogganic contour $x^{(N)}(s)$ into its rectified equivalent
$z^{(0)}(s)$ living inside a single Riemann sheet~${\cal R}_0$.

The situation becomes perceivably less trivial for the quantum
toboggans with a pair of the spires at $x^{(BP)}_{(\pm)}=\pm 1$ in
the QT Schr\"{o}dinger equation of the form (\ref{inio}). In its
combination with a multisheeted contour $x^{(\varrho_0)}$, a
source of simplification will again be sought in an appropriate
change of variables based on the most elementary implicit formula
 \ben
 1+(ix)^2 = \left [
 1+(iz)^2\right ]^\kappa\,,\ \ \ \ \ \ \ \ \kappa > 1\,.
 \een
This combines the conservation of the ${\cal PT}$ symmetry with
the preservation of the position of our pair of branch points.
Another requirement is that our recipe maps the negative imaginary
axis of $z=-{\rm i}\,\varrho$ on itself. This gives the more
explicit candidate for the mapping,
 \be
 x = -{\rm i}\,\sqrt{(1-z^2)^\kappa - 1}\,.
 \label{change}
 \ee
We checked that this recipe deforms smoothly the vicinity of the
negative imaginary axis  at the small angles,
 \ben
 z=-{\rm i}\,{r}\,e^{{\rm i}\,\theta} \ \longrightarrow\ \
 x=-{\rm i}\,
 \left [
 \left (1+{r}^2\,e^{2\,{\rm i}\,\theta}\right )^\kappa-1
 \right ]^{1/2}\,.
 \een
In a way which differs from the single-branch-point recipe of
ref.~\cite{I}, we get just a re-scaling of the coordinate by the
constant factor $\sqrt{\kappa}$ at the small radii ${r}$. Still,
the crucial parallelism taking place at the very large ${r} \gg 1$
is completely preserved. We may conclude that eq.~(\ref{change})
could really represent the necessary rectification recipe for
certain toboggans with the properties which can be deduced from
the more detailed study of eq.~(\ref{change}) at the finite ${r}$.


\begin{figure}[t]                     
\begin{center}                         
\epsfig{file=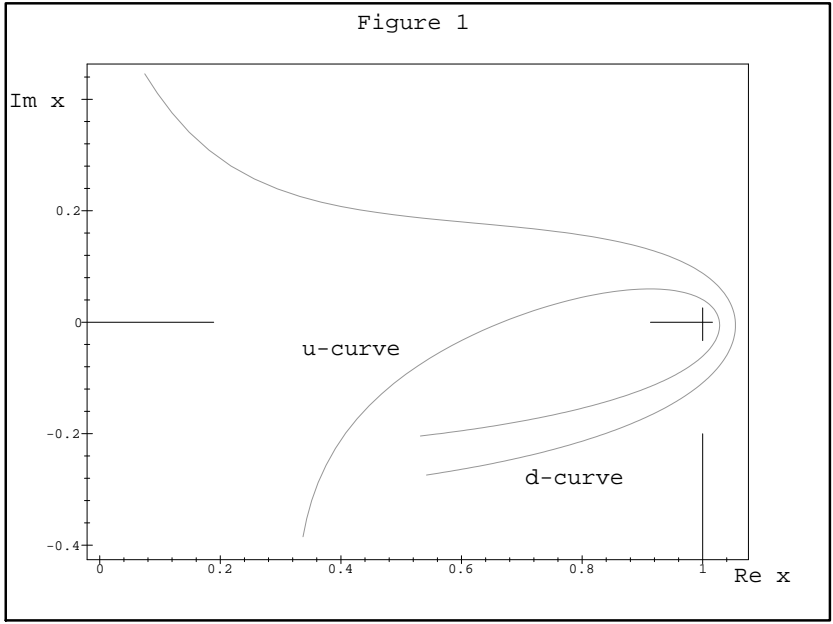,angle=270,width=0.8\textwidth}
\end{center}                         
\vspace{-2mm} \caption{Two samples of the tobogganic curves $x(s)$
obtained as maps of the straight line  of $z$ [parameters
$\kappa=12/5$ and $\varepsilon=3/20$ and $4/20$ were used in
eq.~(\ref{change})].
 \label{obr1}}
\end{figure}

Within the space of this letter, an explicit construction of some
rectifiable samples of the tobogganic paths  can be formulated as
our main numerical task. Its solution starts from the choice of
the straight-line $z(s)=s-{\rm i}\,\varepsilon $ in the rectified,
auxiliary but, presumably, solvable equation (\ref{fino}). We
assume that the change of variables (\ref{change}) is implemented
here in such a manner that it returns us strictly back to the
input, tobogganic bound-state eigenvalue problem of
eq.~(\ref{inio}).

The knot-like structure of the curves $x^{\varrho_0}(s)$ is most
easily visualized when one proceeds from a straight-line $z
(s)=s-{\rm i}\,\varepsilon$. Using the standard computer graphics
facilities, we can only vary the exponent $\kappa>1$ and employ
the mapping ${\cal M}: z(s) \to x(s)$ given by eq.~(\ref{change}).
Such an inverse transformation maps the straight lines $z(s)$
backwards into the equivalent, $\varepsilon-$dependent tobogganic
contours ${x}^{(\varrho_0)}(s)$ where the structure of the
descriptor word $\varrho_0$ is to be inferred from a detailed
analysis of the graphs and pictures.


\begin{figure}[t]                     
\begin{center}                         
\epsfig{file=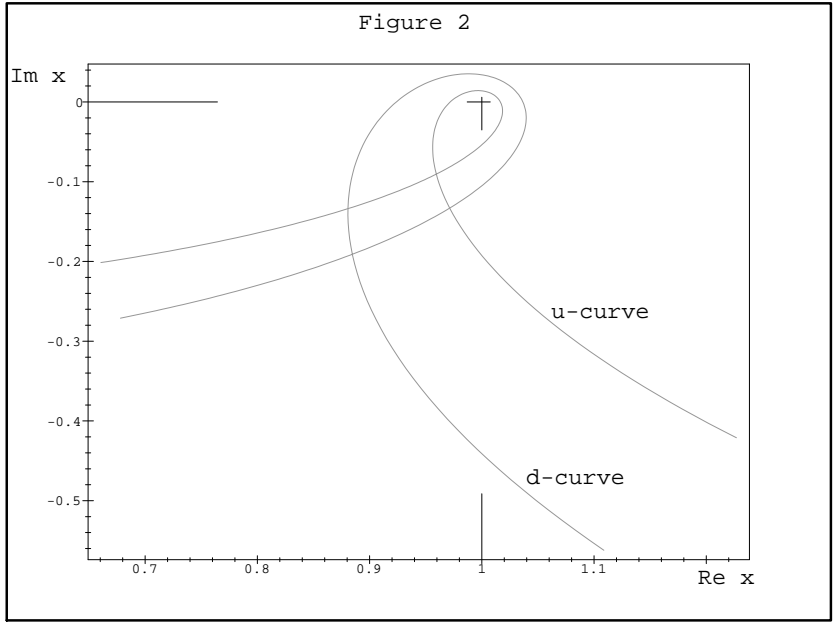,angle=270,width=0.8\textwidth}
\end{center}                         
\vspace{-2mm} \caption{Similar curves as in Figure 1, at the
exponent $\kappa = 3$.
 \label{obr2}}
\end{figure}

In the illustrative Figure~1 we choose $\kappa=2.4$ and restricted
the parameter $s$ to a finite interval between $s=0.4$ and $s=
1.4$. We also selected the two (viz., up- and down-lying)
``representative" samples of the parameter
$\varepsilon=\varepsilon_u=0.15$ and
$\varepsilon=\varepsilon_d=0.2$. This enabled us to illustrate
that and how the explicit shapes and qualitative features of the
resulting curves ${x}^{(\varrho_0)}(s)$ can vary. In order to
guide the eye, we also emphasized the location of the branch point
$x^{(BP)}=+1$.



When we choose a bigger exponent $\kappa=3$, we accelerate the
winding so that our tobogganic spirals $x(s)$ move more quickly
downwards and to the right (cf. Figure~2).


\begin{figure}[t]                     
\begin{center}                         
\epsfig{file=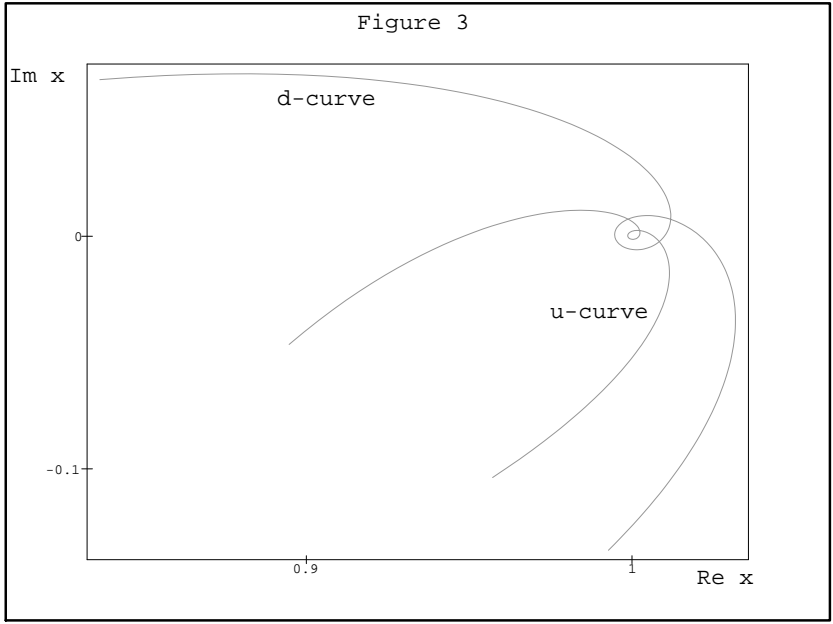,angle=270,width=0.8\textwidth}
\end{center}                         
\vspace{-2mm} \caption{Similar curves as in Figures 1 and 2, at
$\kappa = 5$.
 \label{obr3}}
\end{figure}


When we further increase the value of the exponent to $\kappa=5$,
both our spirals will turn twice around the singularities $\pm 1$
and, with the further growth of $s$, they will continue moving
downwards and to the right in complex plane. In Figure~3 we choose
$s \in (0.55,1.285)$ to show, in more detail, how both our spirals
``orbit" around the point $x^{(BP)}_+=1$.

\subsection{Effective non-tobogganic ${\cal PT}-$symmetric
potentials}

The most important consequence of the preceding graphical
exercises is twofold. Firstly we see that they represent the
easiest way of the determination of the rectifiable descriptors
$\varrho_0$. Secondly,  the closed and compact formulae are at our
disposal. The latter merit of our recipe enables us to
differentiate eq.~(\ref{change}), yielding the relation between
the two differential operators
 \ben
 \frac{d}{d\,x}= \beta(z)\, \frac{d}{d\,z}\,,\ \ \ \ \ \
 \beta(z)=-{\rm i}\,\frac{\sqrt{(1-z^2)^\kappa
 - 1}}{\kappa\,z \,(1-z^2)^{\kappa - 1}}
 \,.
 \een
Together with  the usual ansatz $\psi(x)=\chi(z)\,\varphi(z)$ and
with an abbreviation for the effective potential $V_{eff}({\rm
i}\,x)=V({\rm i}\,x)+2\,\ell(\ell+1)[1-({\rm i}\,x)^2]/[1+({\rm
i}\,x)^2]^2$, the latter formula may be inserted in our
fundamental differential Schr\"{o}dinger eq.~(\ref{inio}) yielding
 \ben
 \left (
 -\beta(z)\,\frac{d}{dz}\,\beta(z)\,\frac{d}{dz}
 +V_{eff}[ix(z)]-E
 \right )\,\chi(z)\,\varphi(z)=0\,.
 \een
The choice of $\chi(z) = const\,/\sqrt{\beta(z)}$ (which dates
back to Liouville \cite{Liouville}) enables us to eliminate the
first derivative of $\varphi(z)$ and to simplify the latter
equation to its final Schr\"{o}dinger form (\ref{fino}). One only
has to add the definition
 \be
 U_{eff}({\rm i}\,z)
 =\frac{V_{eff}[{\rm i}\,x(z)]-E_n}{\beta^2(z)}
 +\frac{\beta''(z)}{2\,\beta(z)}
 -\frac{[\beta'(z)]^2}{4\,\beta^2(z)}
 \label{last}
 \ee
where the prime denotes the differentiation with respect to $z$.
Our task is completed since the ${\cal PT}-$symmetry of the latter
formula is obvious. By the way, it is also amusing to notice that
in the limit of the large $x$ and/or $z$,  our final formula
(\ref{last}) degenerates to its single-spire predecessor of refs.
\cite{I,II} exemplified also here, in paragraph \ref{uho}, by its
sextic-oscillator illustration (\ref{SEHOspiprime}).

\section{Conclusions \label{s5} }

We saw that the introduction of quantum toboggans may be
understood as a mere slight innovation of the traditional
quantization recipes \cite{Carl,Geyer}. Still, in our present
extension of the results of papers \cite{I,II} we demonstrated
that this innovation may still lead to certain very nontrivial
consequences, especially in the context of building
phenomenological models in Quantum Mechanics or even beyond its
scope like, say, in classical magnetohydrodynamics \cite{ZU} etc.

Our present main result is that for each individual potential
$V(x)$ {\em and} for each individual choice of the specific,
sufficiently simple topology (i.e., descriptor
$\varrho=\varrho_0$) of the QT2 path $x^{(\varrho_0)}$, the
necessary proof of the reality/observability of the energies
becomes trivial, based on the mere change of variables given by
closed formulae. Thus, the ``measurable predictions" (i.e., the
bound-state energies) of our $\varrho=\varrho_0$ subset of the QT2
models prove observable {\em if and only if} the ${\cal
PT}-$symmetry of the equivalent rectified problem (\ref{fino})
remains unbroken \cite{Carl}.

In a broader context with any descriptor $\varrho$ we showed that
the ${\cal PT}-$symmetric quantum toboggans with the left-right
symmetric doublet of branching points can be enumerated and
classified using a certain ``word" form of the generalization of
the winding number as used in the most elementary model of
ref.~\cite{I}.

We also discussed in some technical detail how one replaces the
tobogganic Schr\"{o}dinger equation (exhibiting a generalized
${\cal PT}-$symmetry as described in \cite{II}) by its equivalent
representation with the usual ${\cal PT}-$symmetry defined in the
cut complex plane. We paid attention to the fact that  the
rectified, non-tobogganic version of our QT2 Schr\"{o}dinger
bound-state problem contains a more complicated form of the
effective potential, which is the price to be paid for its
simplified definition along an elementary straight line in the
complex plane of $z(s)$.

Obviously, every successful and, in particular, analytic
rectification of ${x^{(QT)}(s)}$ (including, of course, it present
concrete samples) always opens immediately a way to the
translation of a given tobogganic bound-state problem into its
standard straight-line avatar. Thus, in a way which represents a
straightforward QT2 extension of the single-spire QT1 recipe of
ref.~\cite{I}, the existence of the present one-parametric and
compact rectification formula (\ref{change}) may be expected to
faciliatate an application of the standard numerical as well as
purely analytic techniques of the explicit construction of the QT2
bound states at $\varrho=\varrho_0$.

{\it Vice versa}, the possibility of proceeding in the opposite
direction (i.e., from a trivial contour $z^{(0)}$ to its
nontrivial QT2 descendant with $\varrho=\varrho_0\neq 0$) is of
particular appeal in mathematics because in the direction from the
straight line to a tobogganic curve, it enables us to generate a
number of the exactly solvable models with the real spectra (i.e.,
of phenomenological interest).

It is pleasing to see the reducibility of the key problem of the
proof of the reality of the spectrum to the mere technicality of
an appropriate change of variables. The study of its pragmatic
computational consequences has been omitted from the present text
but it may prove equally important in applications. Indeed, one
could find the equivalence between some tobogganic and
non-tobogganic models particularly appealing in the context of
physics.

In many topologically nontrivial Schr\"{o}dinger eigenvalue
problems, the rectification, whenever feasible, would simplify
significantly the explicit and detailed computations of the
topology-dependent spectra. As long as these computations might be
fairly difficult, several detailed numerical studies of this type
(paying attention to the most elementary interaction potentials
$V(x)$) are already under preparation at present
\cite{Milos,Jacek}.

\newpage

\section*{Figure captions}

\subsection*{Figure 1. Two samples of the tobogganic curves $x(s)$
obtained as maps of the straight line  of $y$ [parameters
$\kappa=12/5$ and $\varepsilon=3/20$ and $4/20$ were used in
eq.~(\ref{change})].
 \label{frgal}}

\subsection*{Figure 2. Similar curves as in Figure 1,
at the exponent $\kappa = 3$. \label{rugal}}

\subsection*{Figure 3. Similar curves as in Figures 1 and 2,
at  $\kappa = 5$. \label{zegal}}

\end{document}